\documentclass[aps,prl,reprint,superscriptaddress]{revtex4-1}

\usepackage{graphicx}

\begin{document}


\title{Non-affinity and fluid-coupled viscoelastic plateau for immersed fiber networks}


\author{David Head}
\email[]{d.head@leeds.ac.uk}
\affiliation{School of Computing, University of Leeds, Leeds LS2 9JT, United Kingdom}

\author{Cornelis Storm}
\affiliation{Department of Applied Physics, Eindhoven University of Technology, The Netherlands}
\affiliation{Institute for Complex Molecular Systems, Eindhoven University of Technology, The Netherlands}


\date{\today}


\begin{abstract}
We employ a matrix-based solver for the linear rheology of fluid-immersed disordered spring networks to reveal four distinct dynamic response regimes. One regime---completely absent in the known vacuum response---exhibits coupled fluid flow and network deformation, with both components responding non-affinely. This regime contains an additional plateau (peak) in the frequency-dependent storage (loss) modulus---features which vanish without full hydrodynamic interactions. The mechanical response of immersed networks such as biopolymers and hydrogels is thus richer than previously established, and offers additional modalities for design and control through fluid interactions.
\end{abstract}

\pacs{aaa}

\maketitle

%
%
{\em Introduction.}---Two-phase systems comprising a percolating macromolecular assembly and an interpenetrating fluid arise frequently in nature, and are often synthesized to realize desirable properties~\cite{Bray2001a,Burdick2011a}. When immersed in a viscous solvent, the long-range momentum transfer mediated by fluid hydrodynamics generates non-local physical interactions between the percolating phase~\cite{Batchelor1967a}.
Neglecting such interactions can significantly worsen agreement between models and rheological experiments, lessening our understanding of the function of natural systems, and obscuring rational design principles for synthetic materials. Thus, hydrodynamic interactions are necessary to correctly predict {\em e.g.} the scaling exponents for dilute polymer solutions (Zimm {\em vs.} Rouse)~\cite{Doi1986a,Rubinstein2003a}, sedimentation rates for spheres and semiflexible polymers~\cite{Chaikin1999a,Llopis2008a}, and alignment and clustering of red blood cells in micro-capillary flow~\cite{McWhirter2009a}.

Fiber networks are a class of material for which the effects of hydrodynamic interactions are not fully established.
Examples of these cross-linked assemblies of slender flexible bodies include paper and felt~\cite{Alava2006a}, the eukaryotic cytoskeleton and extra-cellular matrix~\cite{Broedersz2011a}, and the broad range of synthetic hydrogels~\cite{Burdick2011a,Tang2011a,Hoffmann2013a,Li2016a,Rizzi2016a}.
One-way coupling to affine fluid flow has been shown to entrain bond-diluted spring networks at high driving frequencies, leading to affine network deformation ({\em i.e.} uniform across all lengths) that would otherwise be non-affine~\cite{Huisman2010a,Yucht2013a,Amuasi2018a}, and hydrodynamic interactions modifies the exponents describing the loss of rigidity of the same networks~\cite{Dennison2016a}. Network-fluid coupling also explains the frequency-dependent cross-over from negative to positive normal stress~\cite{deCagny2016a,Vahabi2018a}. However, the rheological consequences of deviations from network and fluid affinity as driving frequency and strength of coupling are varied have not been systematically studied, in particular for densities far above the rigidity transition that are relevant to most natural and synthetic systems. A satisfactory understanding of the effects of these couplings is desirable not only from a fundamental perspective, but is also vital to current experimental efforts to design aqueous fiber/polymer materials such as hydrogels to target specific mechanical performance. Cells, for instance, are keenly aware of both the elastic \cite{trappmann2012,engler2006} and the viscous \cite{chaudhuri2015,chaudhuri2016} properties of their substrates, and the ability to rationally engineer materials with tunable properties may help tap into these sensory capacities to purposely elicit different cellular responses.  

\begin{figure}[htbp]
	\centerline{
		\includegraphics[width=8cm]{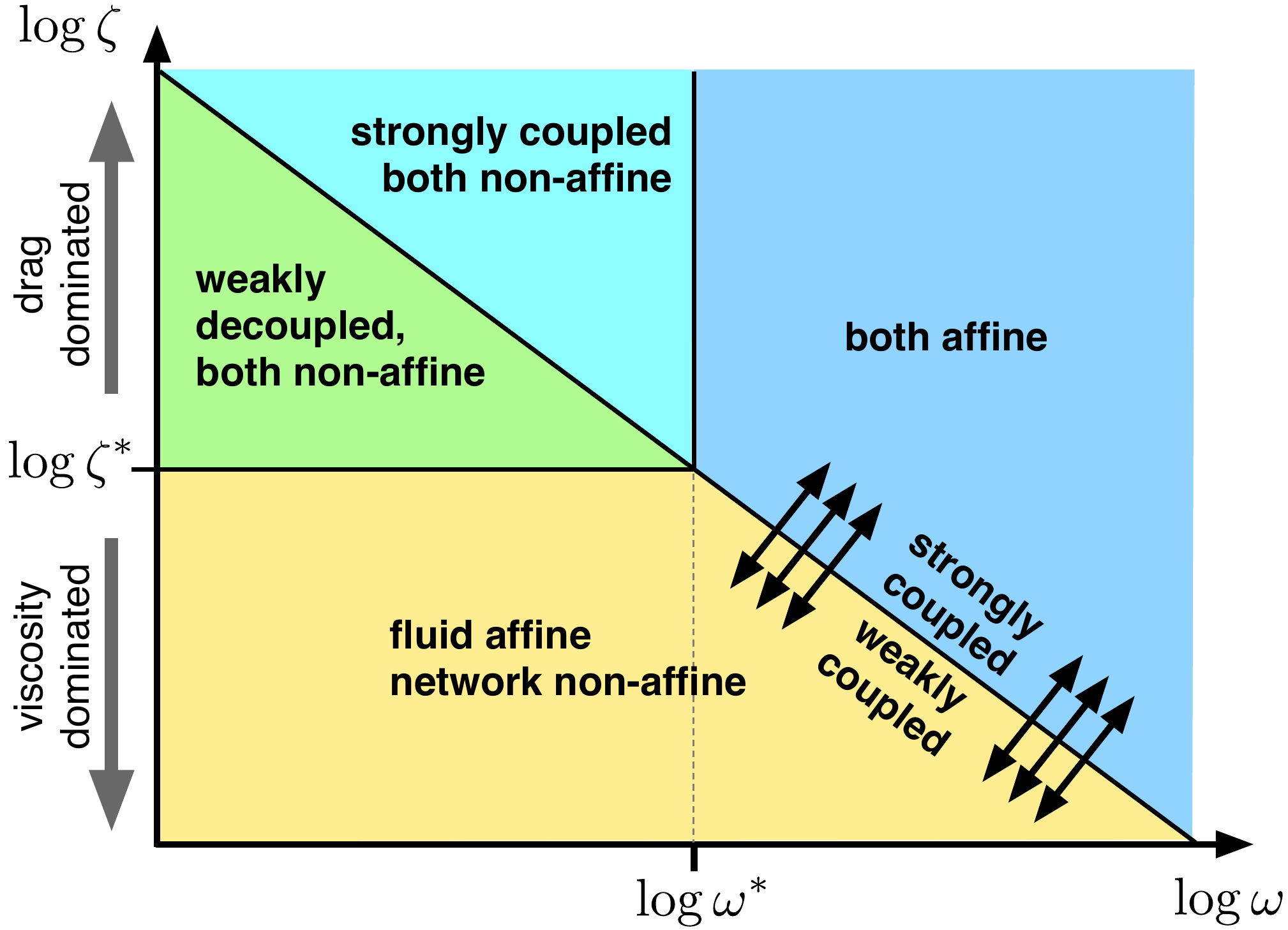}
	}
	\caption{Schematic diagram delineating affine and non-affine response regimes for drag $\zeta$ and driving frequency $\omega$. The `coupled both non-affine' regime is new to this study. The crossover $\zeta^{*}\propto\eta$ with $\eta$ the fluid viscosity, and $\omega^{*}\propto G_{\rm char}/\eta$ with $G_{\rm char}$ some characteristic network stiffness in the absence of fluid. The diagonal line between strongly and weakly coupled response approximately obeys $\zeta\propto\omega^{-1}$.}
	\label{f:responseModes}
\end{figure}

Here we describe an efficient numerical scheme that determines the steady state linear oscillatory response of athermal disordered spring networks immersed in a Stokes fluid.
We quantify frequency ranges for which the network and fluid are weakly or strongly coupled, as a function of network stiffness, fluid viscosity, and the drag coefficient coupling the two. When weakly coupled, the network deforms non-affinely, inducing non-affine fluid flow for high drag but letting the fluid flow affinely when the drag is low. For strong coupling and high frequencies, viscosity dominates and the fluid flows affinely, constraining the network to deform similarly. However, for high drag we also identify a novel coupled regime in which both fluid and network responses are non-affine. This regime, which lies within an arbitrarily-broad, extended frequency range that we identify, exhibits a plateau in the viscoelastic storage modulus intermediate between the low and high-frequency limits, with a peak in the loss modulus at each extreme of the range. Metrics for affinity and coupling are presented that are consistent with the extended nature of this regime.

\begin{figure}[htpb]
	\centerline{
		\includegraphics[width=9cm]{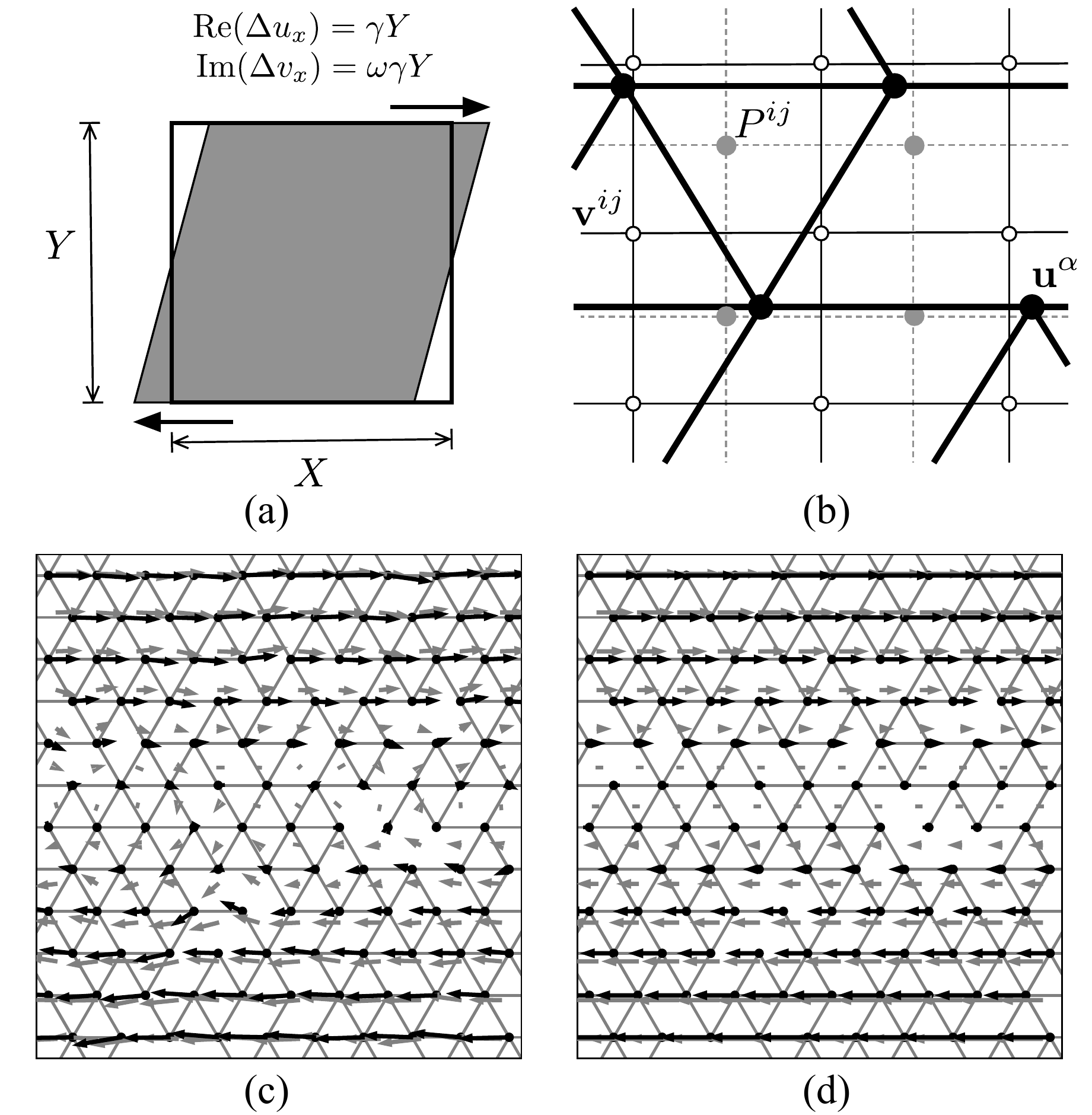}
	}
	\caption{(a)~$X\times Y$ rectangular box with Lees-Edwards boundary conditions mapped to complex amplitudes (see text). (b)~Disordered triangular spring network and fluid meshes, with network node displacements ${\bf u}^{\alpha}$, fluid velocities ${\bf v}^{ij}$, and pressures $P^{ij}$ on a staggered mesh. (c) Solution for a $10\times 12$ network with bond occupation $p=0.8$, drag coefficient $\zeta/\eta=10^{3}$, and driving frequency $\omega\eta/k=10^{-3}$. Black arrows denote in-phase (real) amplitude of network node displacements, and gray arrows denote out-of-phase (imaginary) amplitude of fluid mesh velocities. (d)~Same network with $\omega\eta/k=10$. The maximum arrow length is arbitrary.}
	\label{f:method}	
\end{figure}

%
%
{\em Methodology.}---Our two-dimensional system follows the two-fluid model of MacKintosh and Levine~\cite{MacKintosh2008a}, with the continuum solid replaced by a bond-diluted triangular spring network~\cite{Yucht2013a,Dennison2016a}. Box and mesh geometries are summarized in Figs.~\ref{f:method}(a) and~(b). The network force at node $\alpha$, ${\bf f}^{\alpha}(t)$, is balanced by drag between the node and the surrounding fluid,
\begin{equation}
	{\bf f}^{\alpha}(t) = \zeta \left[
		\partial_{t}{\bf u}^{\alpha}(t) - {\bf v}^{\alpha}(t)
	\right]\:,
	\label{e:drag}
\end{equation}
with $\zeta$ the drag coefficient, ${\bf u}^{\alpha}(t)$ the node displacement and ${\bf v}({\bf x}^{\alpha})$ the fluid velocity at node position ${\bf x}^{\alpha}$. For small displacements and Hookean springs of stiffness $k$,
\begin{equation}
	{\bf f}^{\alpha}(t)
	=
	k
	\sum_{\beta\in N(\alpha)}
	\left\{
		\left[
			{\bf u}^{\beta}(t) - {\bf u}^{\alpha}(t)
		\right]
		\cdot
		\hat{\bf t}^{\alpha\beta}
	\right\}
	\hat{\bf t}^{\alpha\beta}
	\:,
	\label{e:network}
\end{equation}
where $\hat{\bf t}^{\alpha\beta}$ is the unit vector from ${\bf x}^{\alpha}$ to ${\bf x}^{\beta}$, and $\beta\in N(\alpha)$ denotes springs connecting $\alpha$ and $\beta$. The undeformed node separation is the natural spring length~$\ell_{0}$, so there is no prestress. Network disorder is incorporated by removing $1-p$ springs at random, giving a coordination number $\langle z\rangle=6p$. The fluid velocity ${\bf v}({\bf x},t)$ and pressure $P({\bf x},t)$ fields obey steady-state Stoke's equations with the network forces appearing as source terms.
\begin{equation}
	{\bf 0} = \eta \nabla^{2}{\bf v}({\bf x},t)  - \nabla P({\bf x},t) + \sum_{\alpha=1}^{N}{\bf f}^{\alpha}(t)\delta({\bf x} - {\bf x}^{\alpha}),
	\label{e:velocity}
\end{equation}
combined with fluid incompressibility $\nabla\cdot{\bf v}=0$. These equations are discretised using central differences onto staggered rectangular meshes of approximate edge length $0.83\ell_{0}$, 
with ${\bf v}^{ij}(t)$ and $P^{ij}(t)$ at mesh nodes~\cite{Anderson1995a}. Incompressibility and insensitivity of our qualitative findings to fluid mesh size was independently confirmed (see Figs.~S1, S2 in~\cite{SuppMat}). The fluid velocity ${\bf v}^{\alpha}(t)$ at mesh nodes is determined by bilinear interpolation from the fluid mesh.

Oscillatory steady state is assumed for all nodes, ${\bf u}^{\alpha}(t)={\bf u}^{\alpha}e^{i\omega t}=(u_{x}^{\alpha},u_{y}^{\alpha})e^{i\omega t}$, and similarly for ${\bf v}^{ij}(t)$ and $P^{ij}(t)$, and the $e^{i\omega t}$ factors dropped to give linear equations for the complex amplitudes $u_{x}^{\alpha}$, $u_{y}^{\alpha}$, $v_{x}^{ij}$, $v_{y}^{ij}$, and $P^{ij}$. An oscillatory shear $\gamma\cos(\omega t)$ is applied in a Lees-Edwards manner~\cite{Allen1987a} by offsetting the real component of $u^{\alpha}_{x}$ by $\gamma Y$ when the interaction crosses the horizontal boundary, and similarly the imaginary component of $v^{ij}_{x}$ by $\omega\gamma Y$. The discretised equations are assembled into the matrix equation $A{\bf U}={\bf B}$, where vector ${\bf U}=(\{{\bf u}^{\alpha}\},\{{\bf v}^{ij}\},\{P^{ij}\})$ consists of all unknowns, matrix $A$ encodes all network-network, network-fluid and fluid-fluid interactions, and the boundary driving is encoded into vector ${\bf B}$. This is inverted using the sparse direct solver SuperLU~\cite{Demmel1999} to determine the linear, steady-state oscillatory solution for each frequency $\omega$ and network realisation. Examples are given in Figs.~\ref{f:method}(c) and (d). To remove hydrodynamic interactions (retaining only the affine solvent drag), ${\bf v}^{\alpha}(t)$ in (\ref{e:drag}) is replaced with its affine prediction ${\bf v}^{\rm aff}$, and the smaller matrix problem with ${\bf U}=(\{{\bf u}^{\alpha}\})$ solved as before.

All quantities are made dimensionless by scaling with $\eta$ and $k$, {\em i.e.} $\omega \eta/k$, $G^{*}/k$ and $\zeta/\eta$ (note these are two-dimensional).
We consider broad ranges of $\zeta$ to highlight trends and universalities in this class of system, and leave consideration of values for specific materials to experts in the respective domains.

\begin{figure}[htbp]
	\centerline{
		\includegraphics[width=9cm]{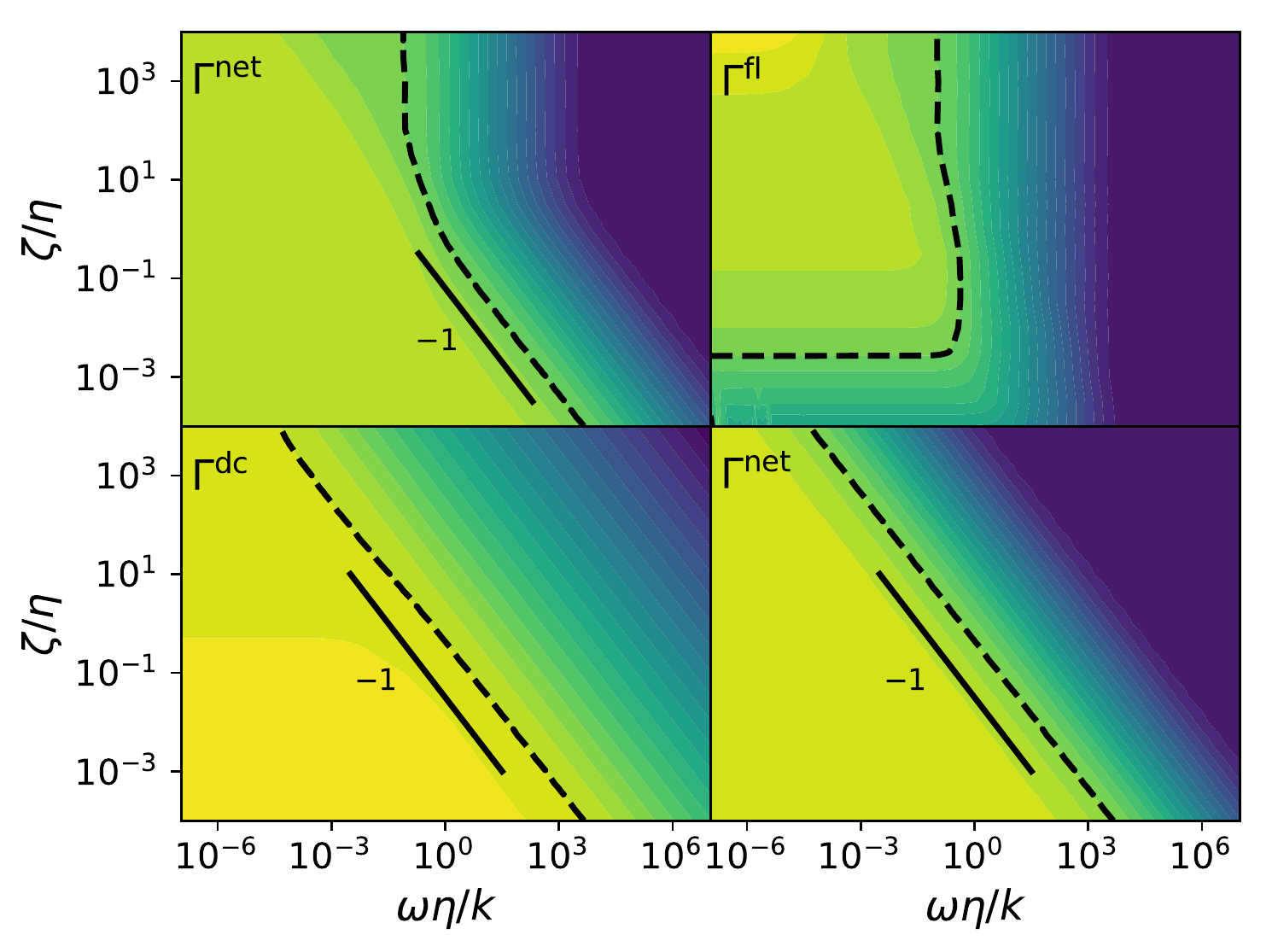}
	}
	\caption{Metrics for network and fluid non-affinity $\Gamma^{\rm net}$ and $\Gamma^{\rm fl}$ and network-fluid decoupling $\Gamma^{\rm dc}$ for $p=0.8$. High (low) values plotted in light (dark), respectively. The lower right-hand panel shows $\Gamma^{\rm net}$ without hydrodynamic interactions. The dashed black lines correspond to a value of 0.1 and the solid black line segments denote lines of constant $\omega\zeta$.}
	\label{f:metrics}
\end{figure}

%
%
{\em Response regimes.}---The degree to which network deformation deviates from affinity can be quantified by generalising the non-affinity metric of~\cite{Broedersz2011a} to complex fields, 
\[
	\Gamma^{\rm net} = \frac{1}{N^{\rm net}\ell_{0}^{2}\gamma^{2}}
	\sum_{\alpha}
	\left\{
		\left| u_{x}^{\alpha} - u_{{\rm aff},x} \right|^{2}
		+
		\left| u_{y}^{\alpha} - u_{{\rm aff},y} \right|^{2}
	\right\},
\]
where ${\bf u}_{\rm aff}=\gamma(y-Y/2,0)$ is the affine prediction and $N^{\rm net}$ the number of network nodes. Additional metrics for fluid non-affinity $\Gamma^{\rm fl}$ and the degree of decoupling between fluid and network $\Gamma^{\rm dc}$ can be similarly defined,
\begin{eqnarray*}
	\Gamma^{\rm fl}
	&=&
	\frac{1}{N^{\rm fl}\omega^{2}\ell_{0}^{2}\gamma^{2}}
	\sum_{ij}
	\left\{
		\left|v_{x}^{ij}-v_{{\rm aff},x}\right|^{2}
		+
		\left|v_{y}^{ij}-v_{{\rm aff},y}\right|^{2}
	\right\},
	\\
	\Gamma^{\rm dc}
	&=&
	\frac{1}{N^{\rm net}\omega^{2}\ell_{0}^{2}\gamma^{2}}
	\sum_{\alpha}
	\left\{
		\left| i\omega u_{x}^{\alpha} - v_{x}^{\alpha} \right|^{2}
		+
		\left| i\omega u_{y}^{\alpha} - v_{y}^{\alpha} \right|^{2}
	\right\},
\end{eqnarray*}
where $N^{\rm fl}$ is the number of fluid mesh nodes, ${\bf v}_{\rm aff}$ denotes affine flow, and $\partial_{t}{\bf u}^{\alpha}(t)\rightarrow i\omega{\bf u}^{\alpha}$. All three dimensionless metrics are plotted in Fig.~\ref{f:metrics} for $p=0.8$, alongside $\Gamma^{\rm net}$ without hydrodynamic interactions. 

The asymptotic response regimes can be inferred by identifying the dominant forces as $\zeta$, $\eta$ and $\omega$ are varied. The magnitude of the drag force (\ref{e:drag}) cannot exceed $\sim \omega\zeta u_{\rm char}$ with $u_{\rm char}$ some characteristic local network displacement, and can be much less than this when $i\omega{\bf u}^{\alpha}\approx{\bf v}^{\alpha}$, {\em i.e.} the network and fluid trajectories coincide. Similarly, the elastic forces (\ref{e:network}) cannot exceed $\sim k u_{\rm char}$, and are much smaller when there is approximate force balance, {\em i.e.} ${\bf u}^{\alpha}\approx{\bf u}^{\alpha}_{0}$ with ${\bf u}_{0}$ the non-affine deformation obeying static equilibrium
(note that James and Guth's prediction of affinity for Gaussian chains does not apply to springs with finite natural length $\ell_{0}>0$~\cite{James1943}).
These considerations suggest that, when $\omega\zeta\gg k$, balance between drag and elastic forces is only possible if the network and fluid move together; we say they are coupled. Similarly, when $\omega\zeta\ll k$ the elastic forces must approach force balance. In the absence of hydrodynamic interactions, this is already enough to infer network affinity for $\omega\zeta/k\gg 1$ and non-affinity for $\omega\zeta/k\ll1$, as confirmed in Fig.~\ref{f:metrics}.

With hydrodynamic interactions, momentum balance must also be obeyed. If $\eta\gg\zeta$, then since the nodal forces ${\bf f}^{\alpha}$ cannot  exceed $\sim\omega\zeta u_{\rm char}$, (\ref{e:velocity}) will be dominated by the viscous term and the fluid will approach the same affine solution ${\bf v}\approx {\bf v}_{\rm aff}$ as in the absence of the network, and $\Gamma^{\rm fl}$ will be low. If in addition $\omega\zeta/k\ll1$, so ${\bf u}^{\alpha}\approx{\bf u}^{\alpha}_{0}$ from above, the network deforms non-affinely ({\em i.e.} $\Gamma^{\rm net}$ is high) and is weakly coupled to the affine fluid flow, {\em i.e.} $\Gamma^{\rm dc}$ will also be high. Conversely, for $\omega\zeta/k\gg1$, the drag tightly couples the network to the affine fluid flow, so the network deforms affinely and $\Gamma^{\rm dc}$ is small. These observations concur with the $\zeta\ll\eta$ range of Fig.~\ref{f:metrics}.

For $\zeta\gg\eta$, three frequency regimes can be identified. If $\omega\zeta/k\ll1$ (so $\omega\eta/k\ll1$ also), ${\bf u}^{\alpha}\approx {\bf u}^{\alpha}_{0}$ and the elastic forces ${\bf f}^{\alpha}$ are expected to be controlled by the drag and of order $\sim\omega\zeta u_{\rm char}$, as in the case without hydrodynamic interactions above. In the absence of significant spatial gradients, the viscous forces $\eta\nabla^{2}{\bf v}$ would scale as $\sim \omega\eta u_{\rm char}$, which cannot balance these ${\bf f}^{\alpha}$. Such gradients must therefore exist, {\em i.e.} the fluid flow is non-affine, but there is no reason to expect ${\bf v}\approx i\omega{\bf u}_{0}^{\alpha}$, so $\Gamma^{\rm dc}$ will be high. This is confirmed in the figure and below, where the viscoelastic spectra are shown to correspond to networks deforming independently of the fluid. In the opposite limit $\omega\eta/k\gg1$ (so $\omega\zeta/k\gg1$ also), drag tightly couples the network to the fluid and momentum balance (\ref{e:velocity}) predicts affine flow, so both network and fluid respond affinely. Intermediate frequencies $\omega\zeta/k\gg1$ and $\omega\eta/k\ll1$ are harder to characterize. Drag dominates network forces leading to tight coupling ${\bf v}^{\alpha}\approx i\omega{\bf u}^{\alpha}$, but no terms in (\ref{e:velocity}) dominate, so a limiting solution for ${\bf v}$ cannot be inferred.
	Reverting to the numerics, Fig.~\ref{f:metrics} suggests a smooth crossover in all metrics from the low to high frequency regimes just identified.

\begin{figure}[htbp]
	\centerline{
		\includegraphics[width=9cm]{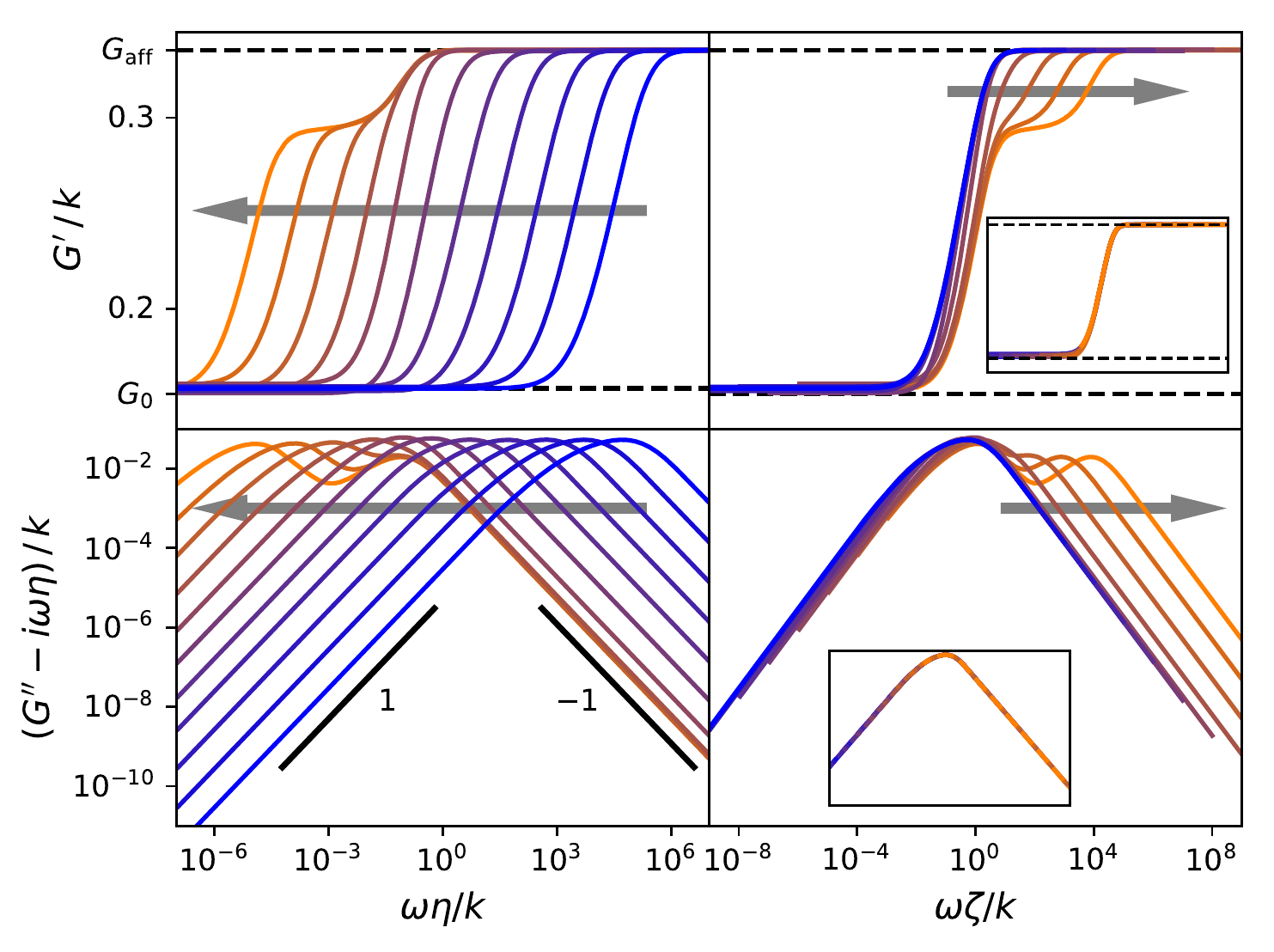}
	}
	\caption{$G^{\prime}(\omega)$ (top panels) and the network contribution to $G^{\prime\prime}(\omega)$ (lower panels) for $p=0.8$, with large arrows showing different $\zeta/\eta$ increasing from $10^{-5}$ (dark curves) to $10^{5}$ (light curves) in factors of ten. The same data is plotted against $\omega\eta/k$ and $\omega\zeta/k$ in the left and right hand panels, resp., and the insets show data over the same ranges without hydrodynamic interactions. The horizontal dashed lines show $G_{\rm aff}=pk\sqrt{3}/4$ and the low-frequency $G_{0}$ evaluated at $\omega\eta/k=10^{-10}$. The line segments have the denoted slope. Each line is averaged over 10 networks of $100\times100$ nodes.}
	\label{f:VESpectra}
\end{figure}

%
%
{\em Viscoelastic spectra.}---The linear viscoelastic response of network plus fluid is quantified by the complex shear modulus $G^{*}(\omega)=G^{\prime}(\omega)+iG^{\prime\prime}(\omega)$, evaluated as~\cite{Yucht2013a}
\begin{equation}
	G^{\star}(\omega)
	=
	\frac{1}{\gamma XY}\sum_{\beta\sim\alpha}f^{\alpha\beta}_{x}r^{\alpha\beta}_{y}
	+
	i\omega\eta\:,
	\label{e:G}
\end{equation}
where ${\bf f}^{\alpha\beta}\equiv{\bf f}^{\beta}-{\bf f}^{\alpha}$ and ${\bf r}^{\alpha\beta}\equiv{\bf x}^{\beta}-{\bf x}^{\alpha}$. 
Note the fluid contribution $i\omega\eta$ does not assume affinity~\cite{SuppMat}.
Results for $p=0.8$ are plotted against both $\omega\eta/k$ and $\omega\zeta/k$ in Fig.~\ref{f:VESpectra}. The previously-reported~\cite{Huisman2010a,Yucht2013a,Rizzi2016a,Dennison2016a,Amuasi2018a} trend for the storage modulus to approach the affine prediction, $G^{\prime}(\omega)\rightarrow G_{\rm aff}=pk\sqrt{3}/4$, as $\omega\rightarrow\infty$, and the static, possibly non-affine limit $G_{0}$  as $\omega\rightarrow 0$, is seen to hold for all $\zeta$. In addition, there is good data collapse when plotted against $\zeta$ when $\zeta\ll\eta$, but this fails when $\zeta\gg\eta$ and hydrodynamic interactions become important. Fig.~S3 in~\cite{SuppMat} shows the same quantities for $p=0.5<p_{\rm c}$, and demonstrates similar scaling but with moduli that vanish as $G^{\prime}(\omega)\sim\omega^{2}$ and $G^{\prime\prime}(\omega)\sim\omega$ as $\omega\rightarrow0$, consistent with the Maxwell model~\cite{Barnes1989}.

\newcommand{\maxZ}{\zeta^{\rm eff}}
\newcommand{\maxE}{\eta^{\rm eff}}
\newcommand{\Gna}{G_{\rm NA}}

For $p>p_{\rm c}$, a plateau emerges in $G^{\prime}(\omega)$ as $\zeta\rightarrow\infty$, lying at a value $\Gna$ between the high and low-frequency plateaus with moduli $G_{\rm aff}$ and $G_{0}$ respectively, and the upper and lower frequencies of this plateau coincide with two peaks in the network contribution to $G^{\prime\prime}(\omega)$. This plateau corresponds to the $\zeta\gg\eta$ intermediate frequency response regime in Fig.~\ref{f:responseModes}, with coupled non-affine fluid and network response. We fit the curves to a spring-dashpot system comprising of a spring and two Maxwell units in parallel, where the Maxwell units have characteristic rates $\maxZ/k$ and $\maxE/k<\maxZ/k$, for which
\begin{eqnarray}
	G^{\prime}(\omega) - G_{0}
	&=&
	(\Gna-G_{0})
	\,g^{\prime}(\omega\maxZ/k)
	\nonumber\\
	&+&
	(G_{\rm aff}-\Gna)
	\,g^{\prime}(\omega\maxE/k)
	\:,
	\\
	G^{\prime\prime}(\omega)
	&=&
	(\Gna-G_{0})
	\,g^{\prime\prime}(\omega\maxZ/k)
	\nonumber\\
	&+&
	(G_{\rm aff}-\Gna)
	\,g^{\prime\prime}(\omega\maxE/k)
	\:,	
\end{eqnarray}
where $g^{\prime}(x)=x^{2}/(1+x^{2})$ and $g^{\prime\prime}(x)=x/(1+x^{2})$.
The upper and lower plateau frequencies extracted from fitting these expressions to the data in Fig.~\ref{f:VESpectra} are given in Fig.~S4 in~\cite{SuppMat}, and show $\maxZ\approx\zeta$ and $\maxE\approx\eta$.

\begin{figure}
	\centerline{
		\includegraphics[width=9cm]{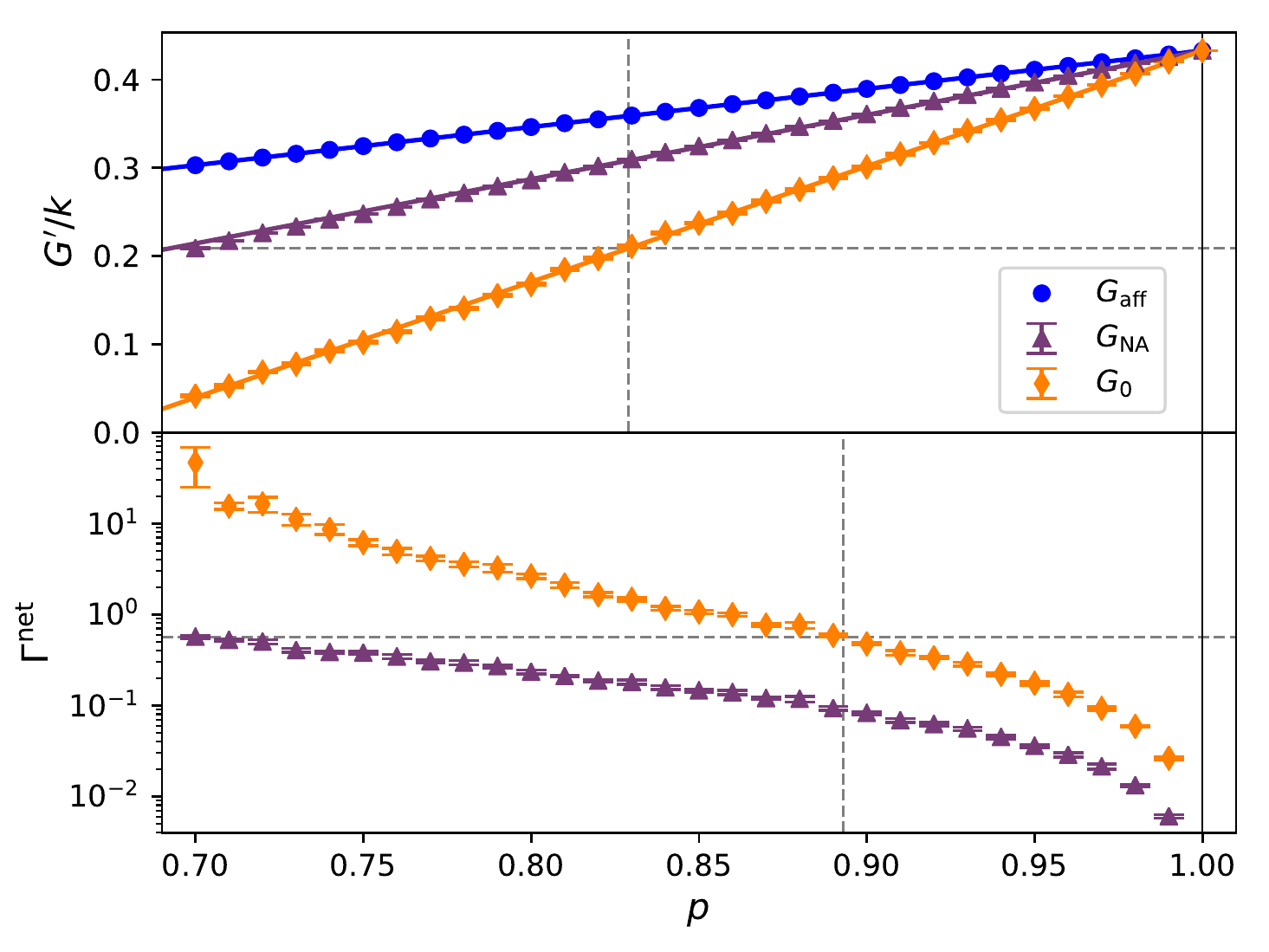}
	}
	\caption{
		{\em (Top)} Shear modulus for the high ($G_{\rm aff}$), intermediate ($G_{\rm NA}$), and low ($G_{0}$) frequency plateaus against $p$, for $\zeta/\eta=10^{5}$. The linear fits are discussed in the text.
		{\em (Bottom)} Network non-affinity $\Gamma^{\rm net}$ for the low-frequency plateau at $\omega\eta/k=10^{-10}$, and the intermediate plateau at $\omega=k/\sqrt{\maxZ\maxE}$, {\em i.e.} midway along the plateau.
		In both panels the dashed lines tie $\Gna$ for $p=0.7$ to $G_{0}$.
	}	
	\label{f:plateaus}
\end{figure}

The plateau values are shown in Fig.~\ref{f:plateaus} alongside corresponding values for $\Gamma^{\rm net}$. All 3 plateau moduli are well described by a linear variation $(k\sqrt{3}/4)(p-p^{*})/(1-p^{*})$ for $p$ sufficiently far above $p_{\rm c}$, with $p^{*}\equiv0$ for $G_{\rm aff}$ by definition of affinity, and fitted values $0.406(2)$ for $\Gna$ and 0.6698(4) for $G_{0}$.
$\Gamma^{\rm net}$ however follows monotonic but non-linear trends, suggesting the systems in the intermediate plateau cannot be mapped to those in the low-frequency plateau for other values of $p$. This is confirmed in the figure, where tie lines from $\Gna$ at $p=0.7$ cannot simultaneously coincide with $G_{0}$ and $\Gamma^{\rm net}$ for any single value of~$p$. We conclude that the non-affine fluid-coupled plateau $\Gna$ deforms in a distinct manner to networks in a vacuum as described by $G_{0}$.

%
%
{\em Discussion.}---
For non-zero damping coefficient $\zeta$ and solvent viscosity $\eta$, we have shown that any finite driving frequency $\omega>0$ rigidifies floppy networks with $p<p_{\rm c}$ (Fig.~S3 in~\cite{SuppMat}). Since most fibrous materials are immersed in a liquid, and $\omega\equiv0$ as stated is impossible to achieve in reality, we can argue that fiber networks should generically be regarded as rigid, albeit possibly very soft. This extends known means to rigidify sub-isostatic networks that includes thermal fluctuations~\cite{Dennison2013}, fibers that resist bending~\cite{Sahimi2003}, and an embedding elastic medium~\cite{vanDoorn2017}. It is not yet clear if known results for $G^{\star}(\omega)$ near $p_{\rm c}$~\cite{Yucht2013a,Dennison2016a} can be expanded into a critical-like `phase' diagram in $(\omega,\zeta,\eta)$-space, similar to these other works. Similarly, it is not known how fluid might affect the strain-induced rigidity transition under continuous (rather than oscillatory) flow~\cite{Vermeulen2017,Merkel2019}.
Investigations into these questions, and improved numerical methodology for immersed fiber networks, would be welcome.
Coupling only at network nodes~\cite{Yucht2013a,Dennison2016a} is numerically convenient, and based on the conceptually-similar bead-spring formalism of the Rouse and Zimm models~\cite{Doi1986a} and rod-based coupling of Huisman {\em et al.}~\cite{Huisman2010a}, which exhibits the same high-frequency limit, we expect equivalent results up to scaling for all coupling schemes. Nonetheless it is desirable to extend more rigorous frameworks~\cite{duRoure2019} to elastic networks to confirm this expectation, in addition to extending this formalism to include semi-flexibility and entanglements.


%
%

\begin{acknowledgments}
	The authors would like to thank Thomas Ranner, Holger Stark and Matthew Dennison for discussions. This work was partly funded by the ECR Internationalisation Activity Fund, University of Leeds, UK. 
\end{acknowledgments}

\bibliography{fibreNetHI}

\begin{thebibliography}{37}%
\makeatletter
\providecommand \@ifxundefined [1]{%
 \@ifx{#1\undefined}
}%
\providecommand \@ifnum [1]{%
 \ifnum #1\expandafter \@firstoftwo
 \else \expandafter \@secondoftwo
 \fi
}%
\providecommand \@ifx [1]{%
 \ifx #1\expandafter \@firstoftwo
 \else \expandafter \@secondoftwo
 \fi
}%
\providecommand \natexlab [1]{#1}%
\providecommand \enquote  [1]{``#1''}%
\providecommand \bibnamefont  [1]{#1}%
\providecommand \bibfnamefont [1]{#1}%
\providecommand \citenamefont [1]{#1}%
\providecommand \href@noop [0]{\@secondoftwo}%
\providecommand \href [0]{\begingroup \@sanitize@url \@href}%
\providecommand \@href[1]{\@@startlink{#1}\@@href}%
\providecommand \@@href[1]{\endgroup#1\@@endlink}%
\providecommand \@sanitize@url [0]{\catcode `\\12\catcode `\$12\catcode
  `\&12\catcode `\#12\catcode `\^12\catcode `\_12\catcode `\%12\relax}%
\providecommand \@@startlink[1]{}%
\providecommand \@@endlink[0]{}%
\providecommand \url  [0]{\begingroup\@sanitize@url \@url }%
\providecommand \@url [1]{\endgroup\@href {#1}{\urlprefix }}%
\providecommand \urlprefix  [0]{URL }%
\providecommand \Eprint [0]{\href }%
\providecommand \doibase [0]{http://dx.doi.org/}%
\providecommand \selectlanguage [0]{\@gobble}%
\providecommand \bibinfo  [0]{\@secondoftwo}%
\providecommand \bibfield  [0]{\@secondoftwo}%
\providecommand \translation [1]{[#1]}%
\providecommand \BibitemOpen [0]{}%
\providecommand \bibitemStop [0]{}%
\providecommand \bibitemNoStop [0]{.\EOS\space}%
\providecommand \EOS [0]{\spacefactor3000\relax}%
\providecommand \BibitemShut  [1]{\csname bibitem#1\endcsname}%
\let\auto@bib@innerbib\@empty
\bibitem [{\citenamefont {Bray}(2001)}]{Bray2001a}%
  \BibitemOpen
  \bibfield  {author} {\bibinfo {author} {\bibfnamefont {D.}~\bibnamefont
  {Bray}},\ }\href@noop {} {\emph {\bibinfo {title} {{Cell movements}}}}\
  (\bibinfo  {publisher} {Garland},\ \bibinfo {address} {New York},\ \bibinfo
  {year} {2001})\BibitemShut {NoStop}%
\bibitem [{\citenamefont {Burdick}\ and\ \citenamefont
  {Mauck}(2011)}]{Burdick2011a}%
  \BibitemOpen
  \bibfield  {author} {\bibinfo {author} {\bibfnamefont {J.~A.}\ \bibnamefont
  {Burdick}}\ and\ \bibinfo {author} {\bibfnamefont {R.~L.}\ \bibnamefont
  {Mauck}},\ }\href@noop {} {\emph {\bibinfo {title} {{Biomaterials for Tissue
  Engineering Applications}}}}\ (\bibinfo  {publisher} {Springer},\ \bibinfo
  {address} {Vienna},\ \bibinfo {year} {2011})\BibitemShut {NoStop}%
\bibitem [{\citenamefont {Batchelor}(1967)}]{Batchelor1967a}%
  \BibitemOpen
  \bibfield  {author} {\bibinfo {author} {\bibfnamefont {G.~K.}\ \bibnamefont
  {Batchelor}},\ }\href@noop {} {\emph {\bibinfo {title} {{An Introduction to
  Fluid Dynamics}}}}\ (\bibinfo  {publisher} {Cambridge University Press},\
  \bibinfo {address} {Cambridge},\ \bibinfo {year} {1967})\BibitemShut
  {NoStop}%
\bibitem [{\citenamefont {Doi}\ and\ \citenamefont {Edwards}(1986)}]{Doi1986a}%
  \BibitemOpen
  \bibfield  {author} {\bibinfo {author} {\bibfnamefont {M.}~\bibnamefont
  {Doi}}\ and\ \bibinfo {author} {\bibfnamefont {S.~F.}\ \bibnamefont
  {Edwards}},\ }\href@noop {} {\emph {\bibinfo {title} {{The Theory of Polymer
  Dynamics}}}}\ (\bibinfo  {publisher} {Oxford University Press},\ \bibinfo
  {address} {Oxford},\ \bibinfo {year} {1986})\BibitemShut {NoStop}%
\bibitem [{\citenamefont {Rubinstein}\ and\ \citenamefont
  {Colby}(2003)}]{Rubinstein2003a}%
  \BibitemOpen
  \bibfield  {author} {\bibinfo {author} {\bibfnamefont {M.}~\bibnamefont
  {Rubinstein}}\ and\ \bibinfo {author} {\bibfnamefont {R.~H.}\ \bibnamefont
  {Colby}},\ }\href@noop {} {\emph {\bibinfo {title} {{Polymer Physics}}}}\
  (\bibinfo  {publisher} {Oxford University Press},\ \bibinfo {address}
  {Oxford},\ \bibinfo {year} {2003})\BibitemShut {NoStop}%
\bibitem [{\citenamefont {Chaikin}(1999)}]{Chaikin1999a}%
  \BibitemOpen
  \bibfield  {author} {\bibinfo {author} {\bibfnamefont {P.}~\bibnamefont
  {Chaikin}},\ }in\ \href@noop {} {\emph {\bibinfo {booktitle} {Soft Fragile
  Matter}}},\ \bibinfo {editor} {edited by\ \bibinfo {editor} {\bibfnamefont
  {M.~E.}\ \bibnamefont {Cates}}\ and\ \bibinfo {editor} {\bibfnamefont
  {M.~R.}\ \bibnamefont {Evans}}}\ (\bibinfo  {publisher} {Institute of
  Physics},\ \bibinfo {address} {Bristol},\ \bibinfo {year} {1999})\ pp.\
  \bibinfo {pages} {315--348}\BibitemShut {NoStop}%
\bibitem [{\citenamefont {Llopis}\ \emph {et~al.}(2008)\citenamefont {Llopis},
  \citenamefont {{Cosentino Lagomarsino}}, \citenamefont {Pagonabarraga},\ and\
  \citenamefont {Lowe}}]{Llopis2008a}%
  \BibitemOpen
  \bibfield  {author} {\bibinfo {author} {\bibfnamefont {I.}~\bibnamefont
  {Llopis}}, \bibinfo {author} {\bibfnamefont {M.}~\bibnamefont {{Cosentino
  Lagomarsino}}}, \bibinfo {author} {\bibfnamefont {I.}~\bibnamefont
  {Pagonabarraga}}, \ and\ \bibinfo {author} {\bibfnamefont {C.~P.}\
  \bibnamefont {Lowe}},\ }\href {\doibase 10.1016/j.cpc.2008.01.014} {\bibfield
   {journal} {\bibinfo  {journal} {Comput. Phys. Commun.}\ }\textbf {\bibinfo
  {volume} {179}},\ \bibinfo {pages} {150} (\bibinfo {year}
  {2008})}\BibitemShut {NoStop}%
\bibitem [{\citenamefont {McWhirter}\ \emph {et~al.}(2009)\citenamefont
  {McWhirter}, \citenamefont {Noguchi},\ and\ \citenamefont
  {Gompper}}]{McWhirter2009a}%
  \BibitemOpen
  \bibfield  {author} {\bibinfo {author} {\bibfnamefont {J.~L.}\ \bibnamefont
  {McWhirter}}, \bibinfo {author} {\bibfnamefont {H.}~\bibnamefont {Noguchi}},
  \ and\ \bibinfo {author} {\bibfnamefont {G.}~\bibnamefont {Gompper}},\ }\href
  {\doibase 10.1073/pnas.0811484106} {\bibfield  {journal} {\bibinfo  {journal}
  {Proc. Natl. Acad. Sci.}\ }\textbf {\bibinfo {volume} {106}},\ \bibinfo
  {pages} {6039} (\bibinfo {year} {2009})}\BibitemShut {NoStop}%
\bibitem [{\citenamefont {Alava}\ and\ \citenamefont
  {Niskanen}(2006)}]{Alava2006a}%
  \BibitemOpen
  \bibfield  {author} {\bibinfo {author} {\bibfnamefont {M.}~\bibnamefont
  {Alava}}\ and\ \bibinfo {author} {\bibfnamefont {K.}~\bibnamefont
  {Niskanen}},\ }\href {\doibase 10.1088/0034-4885/69/3/R03} {\bibfield
  {journal} {\bibinfo  {journal} {Reports Prog. Phys.}\ }\textbf {\bibinfo
  {volume} {69}},\ \bibinfo {pages} {669} (\bibinfo {year} {2006})}\BibitemShut
  {NoStop}%
\bibitem [{\citenamefont {Broedersz}\ \emph {et~al.}(2011)\citenamefont
  {Broedersz}, \citenamefont {Mao}, \citenamefont {Lubensky},\ and\
  \citenamefont {Mackintosh}}]{Broedersz2011a}%
  \BibitemOpen
  \bibfield  {author} {\bibinfo {author} {\bibfnamefont {C.~P.}\ \bibnamefont
  {Broedersz}}, \bibinfo {author} {\bibfnamefont {X.}~\bibnamefont {Mao}},
  \bibinfo {author} {\bibfnamefont {T.~C.}\ \bibnamefont {Lubensky}}, \ and\
  \bibinfo {author} {\bibfnamefont {F.~C.}\ \bibnamefont {Mackintosh}},\ }\href
  {\doibase 10.1038/nphys2127} {\bibfield  {journal} {\bibinfo  {journal} {Nat.
  Phys.}\ }\textbf {\bibinfo {volume} {7}},\ \bibinfo {pages} {983} (\bibinfo
  {year} {2011})}\BibitemShut {NoStop}%
\bibitem [{\citenamefont {Tang}\ \emph {et~al.}(2011)\citenamefont {Tang},
  \citenamefont {Ulijn},\ and\ \citenamefont {Saiani}}]{Tang2011a}%
  \BibitemOpen
  \bibfield  {author} {\bibinfo {author} {\bibfnamefont {C.}~\bibnamefont
  {Tang}}, \bibinfo {author} {\bibfnamefont {R.~V.}\ \bibnamefont {Ulijn}}, \
  and\ \bibinfo {author} {\bibfnamefont {A.}~\bibnamefont {Saiani}},\
  }\href@noop {} {\bibfield  {journal} {\bibinfo  {journal} {Langmuir}\
  }\textbf {\bibinfo {volume} {27}},\ \bibinfo {pages} {14438} (\bibinfo {year}
  {2011})}\BibitemShut {NoStop}%
\bibitem [{\citenamefont {Hoffmann}\ \emph {et~al.}(2013)\citenamefont
  {Hoffmann}, \citenamefont {Tych}, \citenamefont {Hughes}, \citenamefont
  {Brockwell},\ and\ \citenamefont {Dougan}}]{Hoffmann2013a}%
  \BibitemOpen
  \bibfield  {author} {\bibinfo {author} {\bibfnamefont {T.}~\bibnamefont
  {Hoffmann}}, \bibinfo {author} {\bibfnamefont {K.~M.}\ \bibnamefont {Tych}},
  \bibinfo {author} {\bibfnamefont {M.~L.}\ \bibnamefont {Hughes}}, \bibinfo
  {author} {\bibfnamefont {D.~J.}\ \bibnamefont {Brockwell}}, \ and\ \bibinfo
  {author} {\bibfnamefont {L.}~\bibnamefont {Dougan}},\ }\href {\doibase
  10.1039/c3cp52142g} {\bibfield  {journal} {\bibinfo  {journal} {Phys. Chem.
  Chem. Phys.}\ }\textbf {\bibinfo {volume} {15}},\ \bibinfo {pages} {15767}
  (\bibinfo {year} {2013})}\BibitemShut {NoStop}%
\bibitem [{\citenamefont {Li}\ \emph {et~al.}(2016)\citenamefont {Li},
  \citenamefont {Kong}, \citenamefont {Laver},\ and\ \citenamefont
  {Liu}}]{Li2016a}%
  \BibitemOpen
  \bibfield  {author} {\bibinfo {author} {\bibfnamefont {H.}~\bibnamefont
  {Li}}, \bibinfo {author} {\bibfnamefont {N.}~\bibnamefont {Kong}}, \bibinfo
  {author} {\bibfnamefont {B.}~\bibnamefont {Laver}}, \ and\ \bibinfo {author}
  {\bibfnamefont {J.}~\bibnamefont {Liu}},\ }\href {\doibase
  10.1002/smll.201502429} {\bibfield  {journal} {\bibinfo  {journal} {Small}\
  }\textbf {\bibinfo {volume} {12}},\ \bibinfo {pages} {973} (\bibinfo {year}
  {2016})}\BibitemShut {NoStop}%
\bibitem [{\citenamefont {Rizzi}\ \emph {et~al.}(2016)\citenamefont {Rizzi},
  \citenamefont {Auer},\ and\ \citenamefont {Head}}]{Rizzi2016a}%
  \BibitemOpen
  \bibfield  {author} {\bibinfo {author} {\bibfnamefont {L.~G.}\ \bibnamefont
  {Rizzi}}, \bibinfo {author} {\bibfnamefont {S.}~\bibnamefont {Auer}}, \ and\
  \bibinfo {author} {\bibfnamefont {D.~A.}\ \bibnamefont {Head}},\ }\href
  {\doibase 10.1039/c6sm00139d} {\bibfield  {journal} {\bibinfo  {journal}
  {Soft Matter}\ }\textbf {\bibinfo {volume} {12}},\ \bibinfo {pages} {4332}
  (\bibinfo {year} {2016})}\BibitemShut {NoStop}%
\bibitem [{\citenamefont {Huisman}\ \emph {et~al.}(2010)\citenamefont
  {Huisman}, \citenamefont {Storm},\ and\ \citenamefont
  {Barkema}}]{Huisman2010a}%
  \BibitemOpen
  \bibfield  {author} {\bibinfo {author} {\bibfnamefont {E.~M.}\ \bibnamefont
  {Huisman}}, \bibinfo {author} {\bibfnamefont {C.}~\bibnamefont {Storm}}, \
  and\ \bibinfo {author} {\bibfnamefont {G.~T.}\ \bibnamefont {Barkema}},\
  }\href {\doibase 10.1103/PhysRevE.82.061902} {\bibfield  {journal} {\bibinfo
  {journal} {Phys. Rev. E - Stat. Nonlinear, Soft Matter Phys.}\ }\textbf
  {\bibinfo {volume} {82}},\ \bibinfo {pages} {061902} (\bibinfo {year}
  {2010})}\BibitemShut {NoStop}%
\bibitem [{\citenamefont {Yucht}\ \emph {et~al.}(2013)\citenamefont {Yucht},
  \citenamefont {Sheinman},\ and\ \citenamefont {Broedersz}}]{Yucht2013a}%
  \BibitemOpen
  \bibfield  {author} {\bibinfo {author} {\bibfnamefont {M.~G.}\ \bibnamefont
  {Yucht}}, \bibinfo {author} {\bibfnamefont {M.}~\bibnamefont {Sheinman}}, \
  and\ \bibinfo {author} {\bibfnamefont {C.~P.}\ \bibnamefont {Broedersz}},\
  }\href {\doibase 10.1039/c3sm50177a} {\bibfield  {journal} {\bibinfo
  {journal} {Soft Matter}\ }\textbf {\bibinfo {volume} {9}},\ \bibinfo {pages}
  {7000} (\bibinfo {year} {2013})}\BibitemShut {NoStop}%
\bibitem [{\citenamefont {Amuasi}\ \emph {et~al.}(2018)\citenamefont {Amuasi},
  \citenamefont {Fischer}, \citenamefont {Zippelius},\ and\ \citenamefont
  {Heussinger}}]{Amuasi2018a}%
  \BibitemOpen
  \bibfield  {author} {\bibinfo {author} {\bibfnamefont {H.~E.}\ \bibnamefont
  {Amuasi}}, \bibinfo {author} {\bibfnamefont {A.}~\bibnamefont {Fischer}},
  \bibinfo {author} {\bibfnamefont {A.}~\bibnamefont {Zippelius}}, \ and\
  \bibinfo {author} {\bibfnamefont {C.}~\bibnamefont {Heussinger}},\ }\href
  {\doibase 10.1063/1.5030169} {\bibfield  {journal} {\bibinfo  {journal} {J.
  Chem. Phys.}\ }\textbf {\bibinfo {volume} {149}},\ \bibinfo {pages} {084902}
  (\bibinfo {year} {2018})}\BibitemShut {NoStop}%
\bibitem [{\citenamefont {Dennison}\ and\ \citenamefont
  {Stark}(2016)}]{Dennison2016a}%
  \BibitemOpen
  \bibfield  {author} {\bibinfo {author} {\bibfnamefont {M.}~\bibnamefont
  {Dennison}}\ and\ \bibinfo {author} {\bibfnamefont {H.}~\bibnamefont
  {Stark}},\ }\href {\doibase 10.1103/PhysRevE.93.022605} {\bibfield  {journal}
  {\bibinfo  {journal} {Phys. Rev. E}\ }\textbf {\bibinfo {volume} {93}},\
  \bibinfo {pages} {022605} (\bibinfo {year} {2016})}\BibitemShut {NoStop}%
\bibitem [{\citenamefont {{De Cagny}}\ \emph {et~al.}(2016)\citenamefont {{De
  Cagny}}, \citenamefont {Vos}, \citenamefont {Vahabi}, \citenamefont
  {Kurniawan}, \citenamefont {Doi}, \citenamefont {Koenderink}, \citenamefont
  {MacKintosh},\ and\ \citenamefont {Bonn}}]{deCagny2016a}%
  \BibitemOpen
  \bibfield  {author} {\bibinfo {author} {\bibfnamefont {H.~C.}\ \bibnamefont
  {{De Cagny}}}, \bibinfo {author} {\bibfnamefont {B.~E.}\ \bibnamefont {Vos}},
  \bibinfo {author} {\bibfnamefont {M.}~\bibnamefont {Vahabi}}, \bibinfo
  {author} {\bibfnamefont {N.~A.}\ \bibnamefont {Kurniawan}}, \bibinfo {author}
  {\bibfnamefont {M.}~\bibnamefont {Doi}}, \bibinfo {author} {\bibfnamefont
  {G.~H.}\ \bibnamefont {Koenderink}}, \bibinfo {author} {\bibfnamefont
  {F.~C.}\ \bibnamefont {MacKintosh}}, \ and\ \bibinfo {author} {\bibfnamefont
  {D.}~\bibnamefont {Bonn}},\ }\href {\doibase 10.1103/PhysRevLett.117.217802}
  {\bibfield  {journal} {\bibinfo  {journal} {Phys. Rev. Lett.}\ }\textbf
  {\bibinfo {volume} {117}},\ \bibinfo {pages} {217802} (\bibinfo {year}
  {2016})}\BibitemShut {NoStop}%
\bibitem [{\citenamefont {Vahabi}\ \emph {et~al.}(2018)\citenamefont {Vahabi},
  \citenamefont {Vos}, \citenamefont {Cagny}, \citenamefont {Bonn},
  \citenamefont {Koenderink},\ and\ \citenamefont {Mackintosh}}]{Vahabi2018a}%
  \BibitemOpen
  \bibfield  {author} {\bibinfo {author} {\bibfnamefont {M.}~\bibnamefont
  {Vahabi}}, \bibinfo {author} {\bibfnamefont {B.~E.}\ \bibnamefont {Vos}},
  \bibinfo {author} {\bibfnamefont {H.~C. G.~D.}\ \bibnamefont {Cagny}},
  \bibinfo {author} {\bibfnamefont {D.}~\bibnamefont {Bonn}}, \bibinfo {author}
  {\bibfnamefont {G.~H.}\ \bibnamefont {Koenderink}}, \ and\ \bibinfo {author}
  {\bibfnamefont {F.~C.}\ \bibnamefont {Mackintosh}},\ }\href {\doibase
  10.1103/PhysRevE.97.032418} {\bibfield  {journal} {\bibinfo  {journal} {Phys.
  Rev. E}\ }\textbf {\bibinfo {volume} {97}},\ \bibinfo {pages} {032418}
  (\bibinfo {year} {2018})}\BibitemShut {NoStop}%
\bibitem [{\citenamefont {Trappmann}(2012)}]{trappmann2012}%
  \BibitemOpen
  \bibfield  {author} {\bibinfo {author} {\bibfnamefont {B.~{\em et al.}.}\
  \bibnamefont {Trappmann}},\ }\href@noop {} {\bibfield  {journal} {\bibinfo
  {journal} {Nature Materials}\ }\textbf {\bibinfo {volume} {11}},\ \bibinfo
  {pages} {642} (\bibinfo {year} {2012})}\BibitemShut {NoStop}%
\bibitem [{\citenamefont {Engler}\ \emph {et~al.}(2006)\citenamefont {Engler},
  \citenamefont {Sen}, \citenamefont {Sweeney},\ and\ \citenamefont
  {Discher}}]{engler2006}%
  \BibitemOpen
  \bibfield  {author} {\bibinfo {author} {\bibfnamefont {A.~J.}\ \bibnamefont
  {Engler}}, \bibinfo {author} {\bibfnamefont {S.}~\bibnamefont {Sen}},
  \bibinfo {author} {\bibfnamefont {H.~L.}\ \bibnamefont {Sweeney}}, \ and\
  \bibinfo {author} {\bibfnamefont {D.~E.}\ \bibnamefont {Discher}},\
  }\href@noop {} {\bibfield  {journal} {\bibinfo  {journal} {Cell}\ }\textbf
  {\bibinfo {volume} {125}},\ \bibinfo {pages} {677} (\bibinfo {year}
  {2006})}\BibitemShut {NoStop}%
\bibitem [{\citenamefont {Chaudhuri}\ \emph {et~al.}(2015)\citenamefont
  {Chaudhuri}, \citenamefont {Gu}, \citenamefont {Darnell}, \citenamefont
  {Klumpers}, \citenamefont {Bencherif}, \citenamefont {Weaver}, \citenamefont
  {Huebsch},\ and\ \citenamefont {Mooney}}]{chaudhuri2015}%
  \BibitemOpen
  \bibfield  {author} {\bibinfo {author} {\bibfnamefont {O.}~\bibnamefont
  {Chaudhuri}}, \bibinfo {author} {\bibfnamefont {L.}~\bibnamefont {Gu}},
  \bibinfo {author} {\bibfnamefont {M.}~\bibnamefont {Darnell}}, \bibinfo
  {author} {\bibfnamefont {D.}~\bibnamefont {Klumpers}}, \bibinfo {author}
  {\bibfnamefont {S.~A.}\ \bibnamefont {Bencherif}}, \bibinfo {author}
  {\bibfnamefont {J.~C.}\ \bibnamefont {Weaver}}, \bibinfo {author}
  {\bibfnamefont {N.}~\bibnamefont {Huebsch}}, \ and\ \bibinfo {author}
  {\bibfnamefont {D.~J.}\ \bibnamefont {Mooney}},\ }\href@noop {} {\bibfield
  {journal} {\bibinfo  {journal} {Nature Comm.}\ }\textbf {\bibinfo {volume}
  {6}},\ \bibinfo {pages} {6365} (\bibinfo {year} {2015})}\BibitemShut
  {NoStop}%
\bibitem [{\citenamefont {Chaudhuri}(2016)}]{chaudhuri2016}%
  \BibitemOpen
  \bibfield  {author} {\bibinfo {author} {\bibfnamefont {O.~e.~a.}\
  \bibnamefont {Chaudhuri}},\ }\href@noop {} {\bibfield  {journal} {\bibinfo
  {journal} {Nature Materials}\ }\textbf {\bibinfo {volume} {15}},\ \bibinfo
  {pages} {326} (\bibinfo {year} {2016})}\BibitemShut {NoStop}%
\bibitem [{\citenamefont {Mackintosh}\ and\ \citenamefont
  {Levine}(2008)}]{MacKintosh2008a}%
  \BibitemOpen
  \bibfield  {author} {\bibinfo {author} {\bibfnamefont {F.~C.}\ \bibnamefont
  {Mackintosh}}\ and\ \bibinfo {author} {\bibfnamefont {A.~J.}\ \bibnamefont
  {Levine}},\ }\href {\doibase 10.1103/PhysRevLett.100.018104} {\bibfield
  {journal} {\bibinfo  {journal} {Phys. Rev. Lett.}\ }\textbf {\bibinfo
  {volume} {100}},\ \bibinfo {pages} {018104} (\bibinfo {year}
  {2008})}\BibitemShut {NoStop}%
\bibitem [{\citenamefont {Anderson}(1995)}]{Anderson1995a}%
  \BibitemOpen
  \bibfield  {author} {\bibinfo {author} {\bibfnamefont {J.~D.}\ \bibnamefont
  {Anderson}},\ }\href@noop {} {\emph {\bibinfo {title} {{Computational fluid
  dynamics}}}}\ (\bibinfo  {publisher} {McGraw-Hill},\ \bibinfo {address} {New
  York},\ \bibinfo {year} {1995})\BibitemShut {NoStop}%
\bibitem [{Sup()}]{SuppMat}%
  \BibitemOpen
  \href@noop {} {}\bibinfo {note} {See Supplemental Material at [URL will by
  inserted by publisher] for additional figures.}\BibitemShut {Stop}%
\bibitem [{\citenamefont {Allen}\ and\ \citenamefont
  {Tildedsley}(1987)}]{Allen1987a}%
  \BibitemOpen
  \bibfield  {author} {\bibinfo {author} {\bibfnamefont {M.~P.}\ \bibnamefont
  {Allen}}\ and\ \bibinfo {author} {\bibfnamefont {D.~J.}\ \bibnamefont
  {Tildedsley}},\ }\href@noop {} {\emph {\bibinfo {title} {{Computer
  Simulations of Liquids}}}}\ (\bibinfo  {publisher} {Clarendon Press},\
  \bibinfo {address} {Oxford},\ \bibinfo {year} {1987})\BibitemShut {NoStop}%
\bibitem [{\citenamefont {Demmel}\ \emph {et~al.}(1999)\citenamefont {Demmel},
  \citenamefont {Eisenstat}, \citenamefont {Gilbert}, \citenamefont {Li},\ and\
  \citenamefont {Liu}}]{Demmel1999}%
  \BibitemOpen
  \bibfield  {author} {\bibinfo {author} {\bibfnamefont {J.~W.}\ \bibnamefont
  {Demmel}}, \bibinfo {author} {\bibfnamefont {S.~C.}\ \bibnamefont
  {Eisenstat}}, \bibinfo {author} {\bibfnamefont {J.~R.}\ \bibnamefont
  {Gilbert}}, \bibinfo {author} {\bibfnamefont {X.~S.}\ \bibnamefont {Li}}, \
  and\ \bibinfo {author} {\bibfnamefont {J.~W.~H.}\ \bibnamefont {Liu}},\
  }\href {\doibase 10.1137/s0895479895291765} {\bibfield  {journal} {\bibinfo
  {journal} {SIAM J. Matrix Anal. Appl.}\ }\textbf {\bibinfo {volume} {20}},\
  \bibinfo {pages} {720} (\bibinfo {year} {1999})}\BibitemShut {NoStop}%
\bibitem [{\citenamefont {James}\ and\ \citenamefont {Guth}(1943)}]{James1943}%
  \BibitemOpen
  \bibfield  {author} {\bibinfo {author} {\bibfnamefont {H.~M.}\ \bibnamefont
  {James}}\ and\ \bibinfo {author} {\bibfnamefont {E.}~\bibnamefont {Guth}},\
  }\href@noop {} {\bibfield  {journal} {\bibinfo  {journal} {J. Chem. Phys.}\
  }\textbf {\bibinfo {volume} {11}},\ \bibinfo {pages} {455} (\bibinfo {year}
  {1943})}\BibitemShut {NoStop}%
\bibitem [{\citenamefont {Barnes}\ \emph {et~al.}(1989)\citenamefont {Barnes},
  \citenamefont {Hutton},\ and\ \citenamefont {Walters}}]{Barnes1989}%
  \BibitemOpen
  \bibfield  {author} {\bibinfo {author} {\bibfnamefont {H.~A.}\ \bibnamefont
  {Barnes}}, \bibinfo {author} {\bibfnamefont {J.~F.}\ \bibnamefont {Hutton}},
  \ and\ \bibinfo {author} {\bibfnamefont {K.}~\bibnamefont {Walters}},\
  }\href@noop {} {\emph {\bibinfo {title} {{An Introduction to Rheology}}}}\
  (\bibinfo  {publisher} {Elsevier},\ \bibinfo {address} {Amsterdam},\ \bibinfo
  {year} {1989})\BibitemShut {NoStop}%
\bibitem [{\citenamefont {Dennison}\ \emph {et~al.}(2013)\citenamefont
  {Dennison}, \citenamefont {Sheinman}, \citenamefont {Storm},\ and\
  \citenamefont {Mackintosh}}]{Dennison2013}%
  \BibitemOpen
  \bibfield  {author} {\bibinfo {author} {\bibfnamefont {M.}~\bibnamefont
  {Dennison}}, \bibinfo {author} {\bibfnamefont {M.}~\bibnamefont {Sheinman}},
  \bibinfo {author} {\bibfnamefont {C.}~\bibnamefont {Storm}}, \ and\ \bibinfo
  {author} {\bibfnamefont {F.~C.}\ \bibnamefont {Mackintosh}},\ }\href
  {\doibase 10.1103/PhysRevLett.111.095503} {\bibfield  {journal} {\bibinfo
  {journal} {Phys. Rev. Lett.}\ }\textbf {\bibinfo {volume} {111}},\ \bibinfo
  {pages} {095503} (\bibinfo {year} {2013})}\BibitemShut {NoStop}%
\bibitem [{\citenamefont {Sahimi}(2003)}]{Sahimi2003}%
  \BibitemOpen
  \bibfield  {author} {\bibinfo {author} {\bibfnamefont {M.}~\bibnamefont
  {Sahimi}},\ }\href@noop {} {\emph {\bibinfo {title} {{Heterogeneous Materials
  I: Linear Transport and Optical Properties}}}}\ (\bibinfo  {publisher}
  {Springer-Verlag},\ \bibinfo {address} {New York},\ \bibinfo {year}
  {2003})\BibitemShut {NoStop}%
\bibitem [{\citenamefont {van Doorn}\ \emph {et~al.}(2017)\citenamefont {van
  Doorn}, \citenamefont {Lageschaar}, \citenamefont {Sprakel},\ and\
  \citenamefont {{van Der Gucht}}}]{vanDoorn2017}%
  \BibitemOpen
  \bibfield  {author} {\bibinfo {author} {\bibfnamefont {J.~M.}\ \bibnamefont
  {van Doorn}}, \bibinfo {author} {\bibfnamefont {L.}~\bibnamefont
  {Lageschaar}}, \bibinfo {author} {\bibfnamefont {J.}~\bibnamefont {Sprakel}},
  \ and\ \bibinfo {author} {\bibfnamefont {J.}~\bibnamefont {{van Der
  Gucht}}},\ }\href {\doibase 10.1103/PhysRevE.95.042503} {\bibfield  {journal}
  {\bibinfo  {journal} {Phys. Rev. E}\ }\textbf {\bibinfo {volume} {95}},\
  \bibinfo {pages} {042503} (\bibinfo {year} {2017})}\BibitemShut {NoStop}%
\bibitem [{\citenamefont {Vermeulen}\ \emph {et~al.}(2017)\citenamefont
  {Vermeulen}, \citenamefont {Bose}, \citenamefont {Storm},\ and\ \citenamefont
  {Ellenbroek}}]{Vermeulen2017}%
  \BibitemOpen
  \bibfield  {author} {\bibinfo {author} {\bibfnamefont {M.~F.}\ \bibnamefont
  {Vermeulen}}, \bibinfo {author} {\bibfnamefont {A.}~\bibnamefont {Bose}},
  \bibinfo {author} {\bibfnamefont {C.}~\bibnamefont {Storm}}, \ and\ \bibinfo
  {author} {\bibfnamefont {W.~G.}\ \bibnamefont {Ellenbroek}},\ }\href
  {\doibase 10.1103/PhysRevE.96.053003} {\bibfield  {journal} {\bibinfo
  {journal} {Phys. Rev. E}\ }\textbf {\bibinfo {volume} {96}},\ \bibinfo
  {pages} {053003} (\bibinfo {year} {2017})}\BibitemShut {NoStop}%
\bibitem [{\citenamefont {Merkel}\ \emph {et~al.}(2019)\citenamefont {Merkel},
  \citenamefont {Baumgarten}, \citenamefont {Tighe},\ and\ \citenamefont
  {Manning}}]{Merkel2019}%
  \BibitemOpen
  \bibfield  {author} {\bibinfo {author} {\bibfnamefont {M.}~\bibnamefont
  {Merkel}}, \bibinfo {author} {\bibfnamefont {K.}~\bibnamefont {Baumgarten}},
  \bibinfo {author} {\bibfnamefont {B.~P.}\ \bibnamefont {Tighe}}, \ and\
  \bibinfo {author} {\bibfnamefont {M.~L.}\ \bibnamefont {Manning}},\ }\href
  {\doibase 10.1073/pnas.1815436116} {\bibfield  {journal} {\bibinfo  {journal}
  {Proc. Natl. Acad. Sci.}\ }\textbf {\bibinfo {volume} {116}},\ \bibinfo
  {pages} {6560} (\bibinfo {year} {2019})}\BibitemShut {NoStop}%
\bibitem [{\citenamefont {du~Roure}\ \emph {et~al.}(2019)\citenamefont
  {du~Roure}, \citenamefont {Lindner}, \citenamefont {Nazockdast},\ and\
  \citenamefont {Shelley}}]{duRoure2019}%
  \BibitemOpen
  \bibfield  {author} {\bibinfo {author} {\bibfnamefont {O.}~\bibnamefont
  {du~Roure}}, \bibinfo {author} {\bibfnamefont {A.}~\bibnamefont {Lindner}},
  \bibinfo {author} {\bibfnamefont {E.~N.}\ \bibnamefont {Nazockdast}}, \ and\
  \bibinfo {author} {\bibfnamefont {M.~J.}\ \bibnamefont {Shelley}},\
  }\href@noop {} {\bibfield  {journal} {\bibinfo  {journal} {Ann. Rev. Fluid
  Mech.}\ }\textbf {\bibinfo {volume} {51}},\ \bibinfo {pages} {539} (\bibinfo
  {year} {2019})}\BibitemShut {NoStop}%
\end{thebibliography}%

\end{document}